\documentclass[a4paper,twocolumn,prb]{revtex4-2}

\usepackage{amsmath,amssymb}
\usepackage{amsfonts}
\usepackage[usenames,dvipsnames]{pstricks}
\usepackage{epsfig}
\usepackage{units}
\usepackage{float}
\usepackage{color}
\usepackage[breaklinks,pdfborder={0 0 0},colorlinks=true,citecolor=blue,urlcolor=blue,linkcolor=blue]{hyperref}
\usepackage{tikz}
\usepackage{enumerate}

\newcommand{\bk}{\mathbf{k}}
\newcommand{\ket}[1]{|#1\rangle}
\newcommand{\bra}[1]{\langle#1|}
\newcommand{\bracket}[2]{\langle#1|#2\rangle}
\newcommand{\Tr}{\mathrm{Tr}}
\newcommand{\pf}{\mathrm{pfaff}}
\makeindex

\begin{document}
\title{Subgap states at ferromagnetic and spiral-ordered magnetic chains in two-dimensional superconductors. II. Topological classification}

\author{C. J. F. Carroll and B. Braunecker}
\affiliation{SUPA, School of Physics and Astronomy, University of St.\ Andrews, North Haugh, St.\ Andrews KY16 9SS, United Kingdom}

\begin{abstract}
We investigate the topological classification of the subgap bands induced in a two-dimensional
superconductor by a densely packed chain of magnetic moments with ferromagnetic or spiral alignments.
The wave functions for these bands are composites of Yu-Shiba-Rusinov-type states and magnetic scattering states
and have a significant spatial extension away from the magnetic moments. We show that this spatial structure prohibits
a straightforward extraction of a Hamiltonian useful for the topological classification.
To address the latter correctly we construct a family of spatially varying topological Hamiltonians
for the subgap bands adapted for the broken translational symmetry caused by the chain.
The spatial dependence in particular captures the transition to
the topologically trivial bulk phase when moving away from the chain by showing how this, necessarily discontinuous,
transition can be understood from an alignment of zeros with poles of Green's functions.
Through the latter the topological Hamiltonians reflect a characteristic found
otherwise primarily in strongly interacting systems.
\end{abstract}

\maketitle


\section{Introduction}\label{sec:intro}

Until comparatively recently the classification of physical phases relied primarily on the
paradigm of spontaneously broken symmetries introduced by Landau. Over the last decades though this scheme
was complemented by the concept of topological phases. In the latter the symmetries are preserved
but locally similar states can have different global properties,
associated for quantum systems typically with some twists in the wave functions that manifest themselves
only when considering the full ensemble of eigenstates.
The preservation of symmetries remains indeed a key feature of the topological phase classification
as it is on the basis of the existence of symmetry protected, gapped states appearing on entrance
to such phases \cite{Chiu2016}. Such protected states have resulted in a significant body of continually evolving
research with broad and novel potential applications including facilitating the possibility of
topological quantum computing \cite{Pachos2012}.

The universality of the symmetry concept allows quite broadly a characterization of the topological properties
to be made in terms of effective Hamiltonians capturing the generic physics in the vicinity of
points in the Brillouin zone that remain invariant under the specific symmetry operations.
Topological phase transitions are characterized there by gap closures and reopenings, for instance by band
inversion upon tuning of some control parameter.
Most prominent is the invariance under time-reversal symmetry, and in combination with chiral and
parity symmetry this has led to the topological classification table known as the ten-fold way \cite{Schnyder2008,Schnyder2009,Kitaev2009,Ryu2010}.

This type of classification is limited to no or weak interactions though, and strong interactions may
lead to additional phases with intriguing properties.
It is a matter of ongoing research to identify and classify such phases where a broader toolkit is required
beyond the symmetry classification of weakly interacting Hamiltonians \cite{GuWen2009, GuWen2012, Kitaev2011}.
One such tool is the classification based upon Green's functions \cite{Volovik,Gurarie2011,Wang2012a,Wang2012b,Wang2012c,Wang2013,Rachel2018},
which is able to replicate the
success of weakly interacting classifications, whilst allowing the possibility of more readily incorporating strongly interacting phases.

An interesting characteristic arising in a clear way from the Green's function based classification is that topological
phase transitions can arise not only through gap closures at high symmetry points. A topological phase
transition is bound to the generation of topological defects in some global property of the wave functions or
the Hamiltonian when probed over the support of the system's spectrum. The appearance or vanishing of defects requires a singular
behaviour. This is conventionally expressed through the gap closing of the Hamiltonian, corresponding
for the Green's functions to a merger of poles. But it is also possible in the absence
of a gap closure by the merging of zeros of the Green's function \cite{Gurarie2011,Volovik}, or the merging
of a zero and a pole.
As the latter is unlikely to occur in the absence of strong interactions it is not ordinarily
considered. Examples of this phenomenon are thus of significant fundamental interest to better understand
the nature of topological phases broadly. One aspect of this paper is to reveal how such an example can be
extracted from a weakly interacting system with a partially broken spatial translation symmetry.
This results from the necessity of reconsidering how to obtain the topological classification in
such a system, which comprises the other results of this paper.

\begin{figure}[t]
	\centering
	\includegraphics[width=\columnwidth]{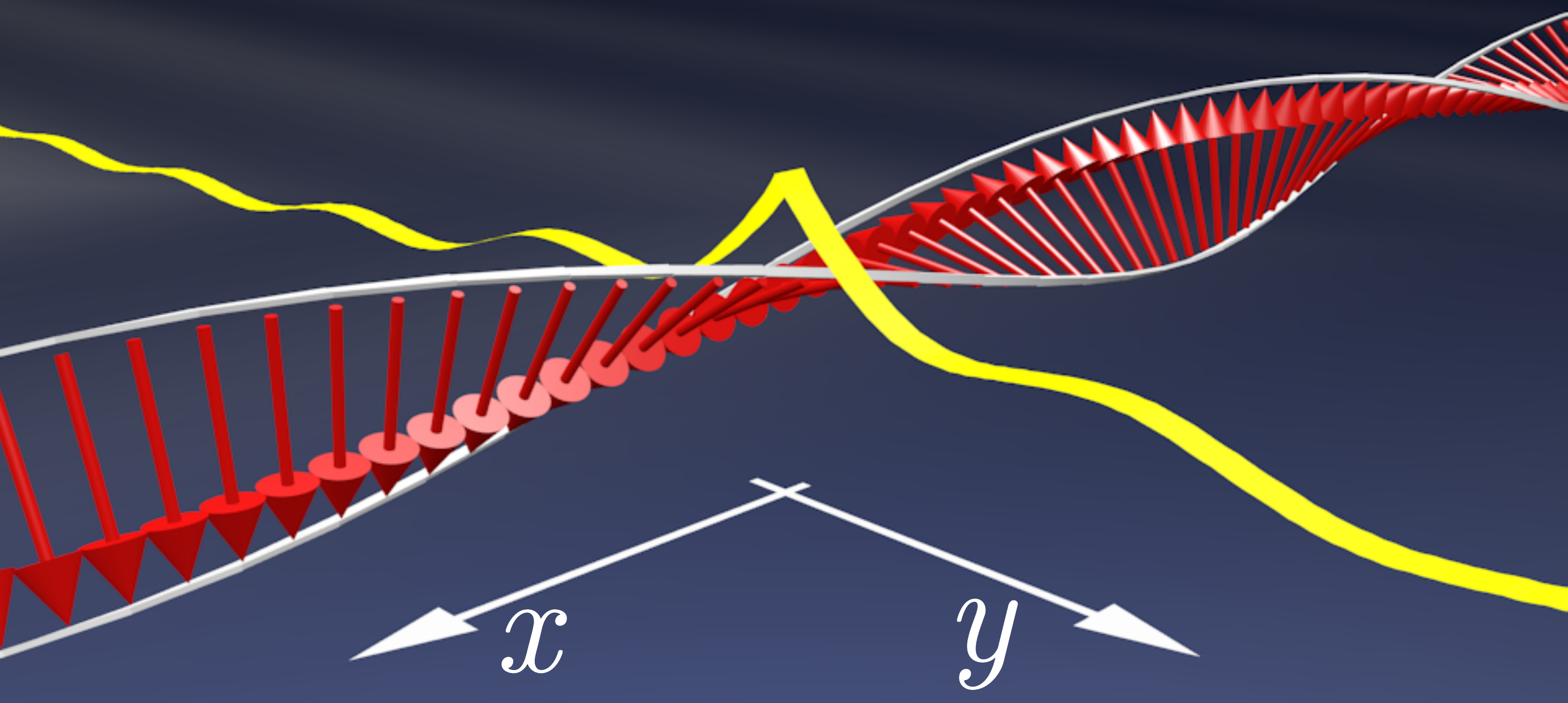}
	\caption{\label{Setup}Schematic representation of the continuous, spiral-ordered line of magnetic moments at $y=0$,
		periodic in $\pi/k_m$ along the $x$ direction. The yellow tape represents the modification of the local
		density of states and thus reflects the spatial extent of the subgap wave functions.}
\end{figure}

Within this work, we build on the model and on key results developed in Ref.\ \cite{PartI}, henceforth called \emph{Part I},
for the system shown in Fig.\ \ref{Setup}, a chain of densely packed magnetic scatterers embedded in a two dimensional (2D)
superconducting substrate. We show that the importance of the spatial structure of the subgap states over all wavelengths
emphasized in Part I has a direct impact on the topological properties too, and we develop a transparent topological
classification which accounts for the lack of translation symmetry.
We indeed demonstrate that although the subgap states are confined near the interface and form
one-dimensional (1D) bands it is not straightforward to eliminate the transverse spatial degree of freedom to
be able to use the established 1D topological classification methods. We in fact provide a rigorous proof
based on Choi's theorem \cite{Choi1975,Stinespring1955} that the often used convenient method of tracing out the
transverse spatial degrees of freedom to obtain an effective 1D Hamiltonian is valid only for fully separable
wave functions. This condition is met, for instance, for confined edge states of quantum Hall systems or topological insulators
for which this elimination is thus applicable. It is, however, not met in the present case, and an uncritical
application of such a method would lead to an incorrect topological classification.

To cure this problem we make use of the full
spatial information of the exact Green's function provided through Part I. The latter comprises in particular the long
spatial extent of the wave functions created from scattering on the magnetic impurities, emphasized earlier
for the long range of Yu-Shiba-Rusinov (YSR) states \cite{Menard2015,Menard2017}.
We introduce a family of spatially varying
topological Hamiltonians that through the standard 1D classification methods provide at the impurity chain
the correct topological invariants, but
also incorporate the transition to the topologically trivial regions of the superconductor at large
distances to the chain. By smoothly varying the distance from the chain, the thus obtained family of topological invariants displays
novel exit and re-entrance into a topologically nontrivial phase due to the interplay between poles and
zeros of the underlying Green's function. This phenomenon occurs in a weakly interacting system and appears to
be entirely due to geometric, interference based considerations. This adds a property to the densely packed
magnetic scatterers that is different to dilute chains of YSR states that can receive a more conventional
1D topological classification which has been amply investigated in the literature
\cite{Choy2011,Kjaergaard2012,NadjPerge2013,vonOppen2013,vonOppen2014,Ojanen2014,Rontynen2014,Lutchyn2014,Kotetes2014,Glazman2014,Franz2014,Heimes2015,Ojanen2015,Brydon2015,Schecter2015,Singh2015,Flensberg2016,Schecter2016,Poyhonen2016,Braunecker2013,Loss2013,Vazifeh2013,Braunecker2015,Peng2015},
starting from the basic phenomenology of YSR states \cite{Yu1965,Shiba1968,Rusinov1969}.
The importance put forward in Part I to determine the exact form of the Green's function of the superconductor with the
magnetic impurity chain becomes essential here.
Indeed we show that only in the domains where the often used long wavelength approximation (LWA) is applicable the use of the
conventional 1D classification methods followed from tracing out the spatial degrees of freedom remains valid.
As the LWA is the extrapolation of tightly packed YSR states this confirms the applicability of the used topological classification.
But it also tells that another approach such as used here is necessary when the LWA no longer applies which, as discussed in Part I,
is in a topologically most interesting range of spiral magnetic order.
Densely packed chains have been realized in experiment and show indeed a more complex band structure
than expected from a simple YSR picture. For such systems the proposed augmented classification method should be directly applicable.

The further structure of the paper is the following.
In Sec.\ \ref{sec:model} we summarize the model and the main result for the Green's function obtained in Part I.
In Sec.\ \ref{sec:1D} we introduce the concept of topological Hamiltonians that will form the basis
for the further discussion. In Sec.\ \ref{sec:comparison_to_1D} we recall the essentials of the topological classification
of the corresponding 1D system. Section \ref{sec:localized_classification} contains the core of this work with the
topological classification tailored to account for the 2D structure of the system. We conclude in Sec.\ \ref{sec:conclusions}.
The analytical results are complemented by a numerical verification based on the tight-binding model
already described in Part I. In the Appendix we discuss the extension to the numerics
for the topological classification.


\section{Model and Green's functions}\label{sec:model}

The model and its properties have been laid out in detail in Part I, and we therefore provide here
only a high-level summary of its main features. We set $\hbar=1$ throughout.
The 2D superconductor is described by the Hamiltonian
\begin{equation} \label{eq:H_0}
	H_0 = \sum_{\mathbf{k},\sigma} \epsilon_{\bk} c^\dagger_{\mathbf{k},\sigma} c_{\bk,\sigma}
	+\bigl( \Delta c_{-\bk,\downarrow}c_{\bk,\uparrow} + \text{h.c.} \bigr).
\end{equation}
Here $c_{\bk,\sigma}$ are the electron operators for momenta $\bk=(k_x,k_y)$ and spins $\sigma=\uparrow,\downarrow = +,-$.
The dispersion $\epsilon_{\bk} = (k_x^2+k_y^2-k_F^2)/2m$ has effective mass $m$ and Fermi momentum $k_F$, and
$\Delta$ is the $s$-wave bulk gap. Spatial coordinates are denoted by $(x,y)$.
The dense chain of classical moments is placed at position $y=0$ and runs along $x$. It scatters electrons
through the Hamiltonian
\begin{equation}  \label{eq:H_m}
	H_m = V_m \int dx \, \mathbf{M}(x) \cdot \mathbf{S}(x,y=0),
\end{equation}
with scattering strength $V_m$,
electron spin operator $\mathbf{S}(x,y)$,
and the planar magnetic spiral formed by the classical spins
$\mathbf{M}(x) = \cos(2 k_m x) \hat{\mathbf{e}}_1+\sin(2 k_m x) \hat{\mathbf{e}}_2$. In the latter expression
the parameter $k_m$
expresses the spiral's periodicity of wavelength $\pi/k_m$ and $\hat{\mathbf{e}}_{1,2}$ are arbitrary orthogonal vectors.
Although self-ordering mechanisms can lead to specific spiral periods \cite{Braunecker2009a, Braunecker2009b,Braunecker2013,Loss2013,Vazifeh2013,Schecter2015,Singh2015,Braunecker2015,Hsu2016}, here
we keep $k_m$ as a free tuning parameter.

The $k_x$ momentum transfer of $2k_m$ by scattering on $H_m$ can be compensated by choosing the spin quantization axis
perpendicular to $\hat{\mathbf{e}}_{1,2}$ and considering the gauge transformation
$c_{\bk,\sigma} \to \tilde{c}_{\bk,\sigma} = c_{(k_x - \sigma k_m, k_y),\sigma}$ \cite{Braunecker2010}.
In this new basis $\mathbf{M}(x) \equiv \hat{\mathbf{e}}_1$ so that $H_m$ corresponds to a ferromagnetic chain of
scattering strength $V_m$ applied perpendicular to the spin quantization axis.
As the transformation also shifts the dispersions $\epsilon_{\bk,\sigma} \to \epsilon_{(k_x+\sigma k_m,k_y)}$
the dispersions of the subgap bands created from scattering on $H_m$ also depend sensitively on $k_m$, and indeed
the spin-dependent shifts are equivalent to a uni-axial spin-orbit interaction \cite{Braunecker2010}.

In the gauge transformed basis translational symmetry along $x$ is restored, and the problem is solved
in a mixed momentum and real space description in the variables $(k_x,y)$. Since $H_m$ induces spin-flip scattering
an extended Nambu-spin basis is required which we choose as
\begin{equation} \label{eq:basis}
	(\tilde{c}^\dagger_{\bk,\uparrow}, \tilde{c}^\dagger_{\bk,\downarrow}, \tilde{c}_{-\bk,\downarrow}, \tilde{c}_{-\bk,\uparrow}),
\end{equation}
with the restriction $k_x\ge 0$ to avoid double counting of states.
Notice that this basis is expressed in the gauge transformed operators,
and does not have the minus sign that is used e.g.\ in front of $\tilde{c}_{-\bk,\uparrow}$ in parts of the literature.
The Pauli matrices acting in Nambu space will be denoted by $\tau_\alpha$ and those acting in spin space by $\sigma_\alpha$,
for $\alpha = x,y,z$. We include furthermore with $\tau_0$ and $\sigma_0$ the corresponding unit matrices.

The system properties are characterized through the retarded Green's function in Nambu-spin space, which for the full
system takes the form
\begin{align}
	G(\omega,k_x,y,y') = \; &g(\omega,k_x,y-y') \nonumber\\
	+ g(\omega,&k_x,y) T(\omega,k_x) g(\omega,k_x,-y'),
\label{eq:G}
\end{align}
where the $T$ matrix is given by the $(\omega_n,k_x)$ dependent matrix
\begin{equation} \label{eq:T}
	T(\omega,k_x) = \bigl[ (V_m \tau_z \sigma_x)^{-1}-g(\omega,k_x,0) \bigr]^{-1},
\end{equation}
and $g(\omega,k_x,y)$ is the bulk Green's function in the absence of $H_m$. For the present model the latter has the
exact solution
\begin{align}
\label{eq:g_k_y}
	&g(\omega,k_x,y)
	=\sum_{\sigma}
	\frac{-i \pi \rho}{2 k_F\sqrt{\tilde{\omega}^2-\tilde{\Delta}^2 + i \eta}}
	\\
	&\times
	\left\{
		\tilde{\omega}_+ \xi_\sigma  \tau_0^{\sigma} + \sigma \tilde{\Delta} \xi_\sigma \tau_x^{\sigma}
		+
		\left[ (\kappa_{\sigma}^2-1) \xi_\sigma +\chi_\sigma \right] \tau_z^{\sigma}
	\right\},
\nonumber
\end{align}
where
$\rho = m / \pi$ is the 2D density of states at the Fermi energy,
$\tilde{\omega}=\omega/E_F$ and $\tilde{\Delta}=\Delta/E_F$ are dimensionless frequency and gap, for $E_F = k_F^2/2m$,
$\tau_{\alpha}^{\pm}=\tau_{\alpha}(\sigma_0 \pm \sigma_z)/2$,
$\eta>0$ is an infinitesimal shift and $\tilde{\omega}_+ = \tilde{\omega}+i\eta$.
Furthermore we have defined
\begin{align}
	\kappa_{\sigma} &= (k_x + \sigma k_m)/k_F,
\\
	\xi_\sigma      &= p_{\sigma,+}^{-1}\mathrm{e}^{i |y|k_F p_{\sigma,+}}+ p_{\sigma,-}^{-1} \mathrm{e}^{-i |y|k_F p_{\sigma,-}},
\\
	\chi_\sigma     &= p_{\sigma,+} \mathrm{e}^{i |y|k_F p_{\sigma,+}}+p_{\sigma,-} \mathrm{e}^{-i |y|k_F p_{\sigma,-}},
\end{align}
with
\begin{equation} \label{eq:p_sigma_pm}
	p_{\sigma,\pm}=\bigl[1-\kappa^2_{\sigma}\pm (\tilde{\omega}^2-\tilde{\Delta}^2 + i\eta)^{1/2}\bigr]^{1/2}.
\end{equation}
In Part I we provided a detailed analysis of the importance of using the Green's function of Eq.\ \eqref{eq:g_k_y} and not
any commonly used approximations. Equation \eqref{eq:g_k_y} remains of fundamental importance in this paper, as any
such approximation would lead to an incorrect topological classification.

The direct computation of $G$ and $T$ consists of a number of
matrix multiplications and inversions and this last step is generally done numerically, though the relatively simple form allows for a number of analytic results which we summarize in the following.

The poles of the Green's function provide the spectrum, and all
subgap states arise from the poles of the $T$ matrix, hence $\det T^{-1}=0$ at some $|\omega|<\Delta$ provides the criterion for
the existence of a subgap state.
The solution of $\det T^{-1}(\omega=0,k_x=0)=0$ is of particular interest because it provides the condition for the interaction strength
$V_m$ at which the subgap states close the gap at the high symmetry point. One can analytically solve this equation for
any spiral wavevector. If we define with
\begin{equation} \label{eqn:C_m}
	C_m = \pi \rho V_m / k_F,
\end{equation}
the dimensionless amplitude of the magnetic scattering strength, then the critical amplitude for the
gap closure is given by
\begin{align}\label{eqn:kx0gapclosure}
	C_m^\star = \bigl[\bigl(1 - k_m^2/k_F^2\bigr)^2 + \tilde{\Delta}^2\bigr]^{1/4}.
\end{align}
As discussed in Part I the exact result of Eq.\ \eqref{eqn:kx0gapclosure} bears a number of interesting features.
The exponent of $1/4$ rather than $1/2$ as expected by comparison to a purely 1D model (see Sec.\ \ref{sec:comparison_to_1D}) occurs due to the dimensional mismatch between the substrate and the impurity chain.
At a ferromagnetic interface with $k_m=0$ the gap closing has only a weak dependence on $\tilde{\Delta}$ and can
be interpreted as the result from the hybridization between the YSR states forming the Shiba bands. On the other hand at $k_m=k_F$
one has $C_m = \tilde{\Delta}^{1/2}$, and thus a gap closure caused by the direct competition between magnetic scattering and pairing.
This resembles a dimensionally renormalized Zeeman interaction, and as shown in Part I indeed YSR states and their hybridization are of no importance in this limit. We additionally point out that Eq.\ \eqref{eqn:kx0gapclosure} is in excellent agreement with the self-consistent numerical solution of the lattice version of this problem, showing that Eq.\ \eqref{eqn:kx0gapclosure} is indeed a general result and not specific to the chosen continuum model.


\section{Topological Hamiltonians}\label{sec:1D}

The topological classification of a material is based on the calculation of topological indices.
Two types of approaches are common for bulk superconductors, based on either characteristics of the Hamiltonian
at special points or integrals of Berry type connections over the Brillouin zone.
In the former category falls the common  $\mathbb{Z}_2$ characterization determined from the sign of
Pfaffians of matrices proportional to the Hamiltonian at time-reversal symmetric points in the
Brillouin zone \cite{Kitaev2001,Kane2005,Fu2006,Stanescu2011}.
With such an approach the classification of the purely 1D system of Sec.\ \ref{sec:comparison_to_1D} is
immediate. The latter category refers to topological indices expressed for example through
TKNN invariants, Chern numbers and Zak phases \cite{TKNN,Zak1989}. These cases
require the knowledge of the Bloch wavefunctions. Equivalent indices can be obtained through Green's
functions \cite{Volovik,Wang2010} which has the advantage that interactions can be included as well
\cite{Gurarie2011,Wang2012a,Wang2012b,Rachel2018}.
Yet in their original formulations these indices involve multiple products of Green's functions,
their derivatives, and frequency integrals in addition to momentum integrals.
A large effort was therefore made to derive simpler equivalent expressions \cite{Wang2012a,Wang2012b,Wang2012c,Wang2013}.
Notable is the replacement of the frequency $\omega$ integral by $\omega = 0$ and use of the Green's function
then to define an effective topological Hamiltonian that correctly captures the topological
classification \cite{Gurarie2011,Wang2012c,Wang2013,Budich2013,Weststrom2016,Xie2020}.
The latter is indeed rather intuitive since any Green's function is obtained through matrix elements
of the resolvent $\hat{G}(\omega) = ( \omega - H )^{-1}$ such that $H = - \hat{G}^{-1}(0)$.
Subtleties arise since Green's functions are projections of the resolvent and their inversion
does not reproduce the original (possibly interacting) Hamiltonian. But, notwithstanding the subtleties, they correctly capture
the topological classification \cite{Wang2012c,Wang2013}.

For a bulk system the topological Hamiltonian can be defined through
\begin{equation} \label{eqn:H_top_bulk}
	H^\text{top}_\text{bulk}(\bk) = - G_\text{bulk}^{-1}(\omega=0,\bk),
\end{equation}
where $G_\text{bulk}$ is the Green's function of the fully translationally symmetric system.

In the following we will show that a similar approach can be adopted for our situation, although
we have neither translational symmetry nor a periodic structure along the $y$ situation.
Despite this, we will demonstrate that a suitably adapted variant of Eq.\ \eqref{eqn:H_top_bulk}
produces the correct topological classification if subtleties with the $y$ dependence are
appropriately taken into account.


\section{Comparison to 1D system}\label{sec:comparison_to_1D}

To obtain a baseline for the expected topological classification we start by providing a brief account
of the straightforward topological classification of a purely 1D model, along with the expected dimensional
renormalization due to the embedding in a 2D system.

The 1D equivalent of Hamiltonian $H=H_0+H_m$ [Eqs.\ \eqref{eq:H_0} and \eqref{eq:H_m}] is in the gauge transformed
basis
\begin{align}\label{eqn:1D_Hamiltonian}
	H(k_x) = \sum_{\sigma} \epsilon_{k_x+\sigma k_m} \tau_z^\sigma + \check{V}_m \tau_z \sigma_x + \Delta \tau_x \sigma_z,
\end{align}
written here not in second quantized form but as a $4\times 4$ matrix in Nambu-spin space at fixed $k_x$.
We identify $\hat{\mathbf{e}}_1$ with the spin-$x$ direction and $\tau_z^\pm = \tau_z(\sigma_0 \pm \sigma_z)/2$
and denote the magnetic potential $\check{V}_m$ to avoid confusion with its counterpart in the 2D system.
Since the $\check{V}_m$ act on the entire system and not only on a line across the 2D system they take the role of a uniform magnetic field
whose original spiral was unwound through the gauge transformation.
Equation \eqref{eqn:1D_Hamiltonian} corresponds to the Hamiltonian of a ``Majorana wire'' \cite{DasSarma2010,vonOppen2010,Lutchyn2010}, which has a known $\mathbb{Z}_2$ topological classification that can be obtained from the Pfaffians of the Hamiltonian at the time-reversal symmetric momenta \cite{Kitaev2001}.

\begin{figure}
	\centering
	\includegraphics[width=\columnwidth]{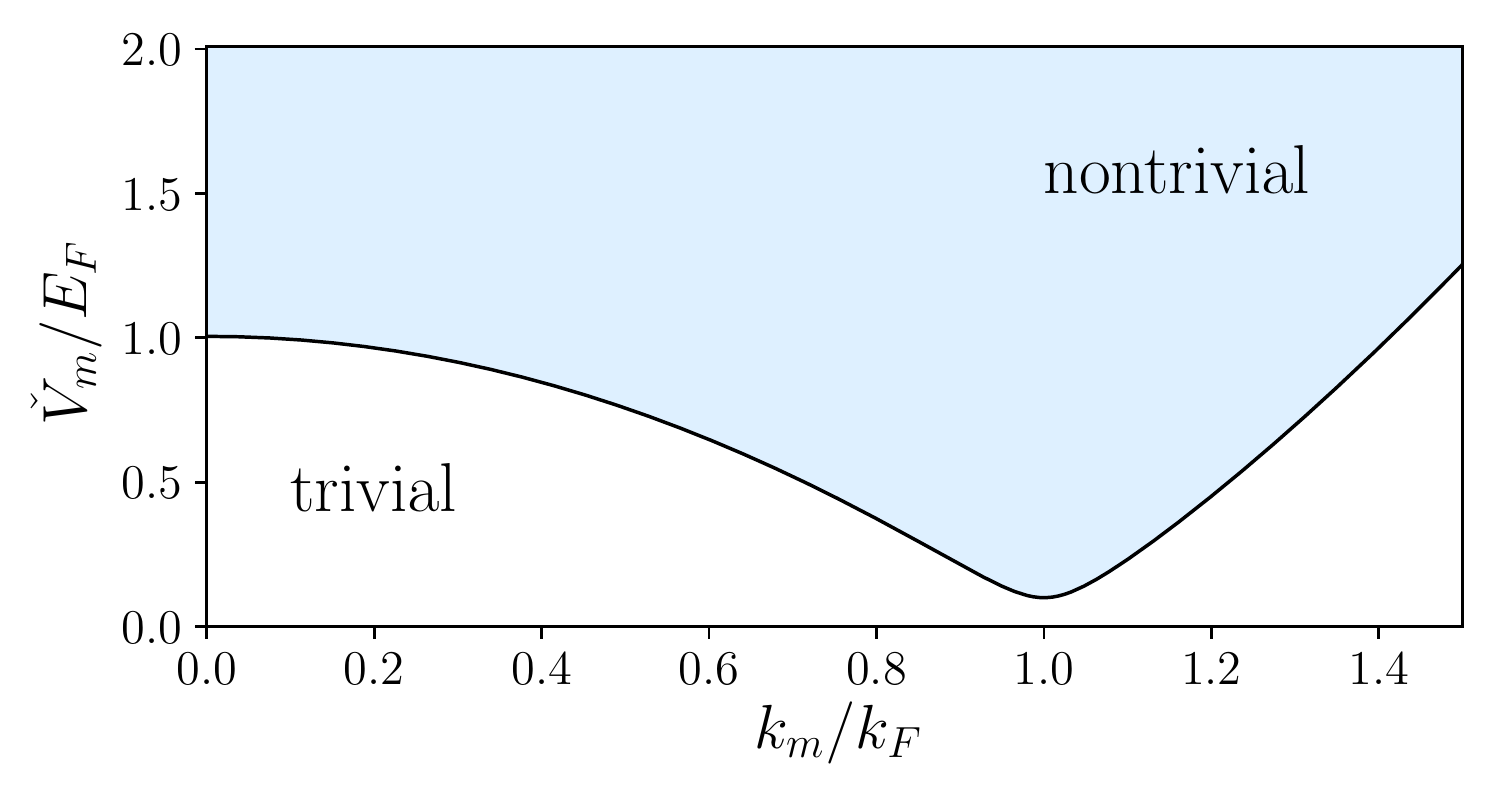}
	\caption{\label{fig:1D_phase_diagram}%
	Topological phase diagram for a pure 1D system with a spiral magnetic field, shown as function
	of field winding momentum $k_m$ versus field strength $\check{V}_m$, for $\Delta = 0.1 E_F$.
	The white are is topologically trivial and the
	shaded area is nontrivial. The separating curve is described by Eq.\ \eqref{eqn:1D_phase_boundary}.
	In this pure 1D case the minimum at $k_m=k_F$ is reached at $\check{V}_m = \Delta$.
	}
\end{figure}

Using this Hamiltonian we calculate the topological invariant in the usual way by transforming $H(0)$ to a skew symmetric matrix $UH(0)$,
where $U=\sigma_x \tau_x$ [taking this form because of the chosen Nambu-spin basis given by Eq.\ \eqref{eq:basis}], and by determining the sign of
the Pfaffian $\pf[UH(0)]$. The resulting phase diagram is plotted in Fig. \ref{fig:1D_phase_diagram} and shows the two distinct topological phases with the transition controlled by $\check{V}_m$. We should remark that for the continuum model there is only one time-reversal symmetric momentum, $k_x=0$, whereas in a lattice system there would also be the momentum at the boundary of the Brillouin zone.
In the latter this second momentum is responsible for a re-entrance to the topologically trivial phase at large magnetic interaction strength
which is absent in the present continuum model.

The boundary between two topologically distinct phases is characterized by a gap closure at a time-reversal invariant momentum.
If we set $\check{C}_m = \check{V}_m/E_F$ in analogy to Eq.\ \eqref{eqn:C_m} and $\check{\Delta}=\Delta/E_F$, the gap closure at $k_x=0$ for the 1D Hamiltonian
requires an interaction strength $\check{C}_m = \check{C}_m^\star$, with
\begin{align}\label{eqn:1D_phase_boundary}
	\check{C}_m^\star = \bigl[\bigr(1- k_m^2/k_F^2\bigl)^2 + \check{\Delta}^2\bigr]^{1/2}.
\end{align}
This critical amplitude has the same functional form as its 2D counterpart $C^\star$ given in Eq.\ \eqref{eqn:kx0gapclosure}
but with the exponent $1/2$ instead of $1/4$. This change is a dimensional renormalization, as mentioned above and explained further in Part I, due to the fact that in contrast to the 2D case $\check{V}_m$ acts on the full transverse extension of the wave functions.
Besides this dimensional renormalization the subgap states remain confined to the vicinity of the magnetic chain. We may thus expect that they retain a 1D character so that up to a renormalization of the phase boundaries the phase diagram itself remains unchanged from the 1D case. As a motivational argument we may indeed consider a procedure that continuously provides an increasing confinement transforming the 2D system into the pure 1D system. If this is done in each gapped phase in a manner such that the gap never closes then the topological class of the subgap states should not change.

Such an argument alone, however, is naive as it neglects that in the 1D case an extra confining potential is required, whereas in 2D the
confinement of the subgap states is controlled by $\Delta$. In the transition between 1D and 2D there is therefore a length scale at which
the boundary condition for the confinement changes its physical origin. Since topology depends on global properties of the wave functions
a change of boundary condition must always be considered carefully, and we will see that indeed the extension of wave functions
across this scale is of importance.
We therefore must consider in more detail the subtleties arising from the loss of translational symmetry or periodicity in the $y$ direction.


\section{Localized classification}\label{sec:localized_classification}

\subsection{Absence of an effective 1D Hamiltonian} \label{sec:absence_1D_H}

Due to the exponential confinement of the subgap wave functions to the region near the magnetic chain the electron motion is one-dimensional.
One may thus consider a description in terms of an effective 1D Hamiltonian, similar to those used for the 1D states appearing through confinement in heterostructures or to the edge bands in topological systems.

A complication arises here from the fact that the Nambu-spin and $y$ degrees of freedom are highly mixed in the wave functions, as visible in Eqs.\ \eqref{eq:g_k_y}--\eqref{eq:p_sigma_pm}, whereas the conventional topological classification tools for 1D systems rely on the Nambu-spin structure alone. In the following sections we provide a systematic discussion that in such a case the topology of the subgap states can be reliably extracted by pinning $y$ to the special value $y=0$, followed by an exploration of the changes for $y \neq 0$.

In this section, however, we analyze the conditions under which the $y$ coordinate can be traced out entirely while maintaining the validity and convenience of the Nambu-spin based classification scheme. We formulate two conditions, (a) and (b) below, that a reduced Hamiltonian $H_\text{eff}^\text{1D}$ should fulfil and show that these conditions have a close connection with Choi's theorem on completely positive trace preserving maps \cite{Choi1975,Stinespring1955}. Based on this we demonstrate that fulfilment of the conditions necessarily imposes a complete separability of the Nambu-spin and $y$ degrees of freedom. This separability is generally not fulfilled in the present case and thus such an effective Hamiltonian cannot be constructed.
A notable exception though is the regime in which the LWA is valid. For the latter the necessary separability is approximately true, explaining why for such a situation a topological classification based on a simple tracing out of $y$ provides correct results.

A dimensional reduction is an often tacitly used procedure in the study of low dimensional systems. A quantum dot, for instance, is addressed commonly by operators creating and annihilating its different levels as entities without addressing the specific spatial structure. Interactions such as the Coulomb repulsion or spin-orbit are effective integral quantities coupling the different levels. Such a description results from
first analyzing the confinement of some non-interacting Hamiltonian, providing the set of basis functions for the confined geometry, and then expanding the full Hamiltonian in this basis. The eigenstates and the spectrum are then obtained by the diagonalization of the resulting Hamiltonian matrix, with the eigenstates given by an appropriate decomposition of the basis functions.

Our situation is distinct in that we already fully know the confined eigenstates. We have thus a different goal with the extraction of a lower-dimensional Hamiltonian. As explained above our goal is to be able to work with the Nambu-spin based symmetries and topological classification methods without having to maintain the $y$ dependence and, especially, without having to modify the methods.

The following proof when this is possible is not specific for the considered situation but general for any type of Hamiltonian with a finite subset of discrete, localized states that are split off from the continuum.

For a fixed $k_x$ any full 2D Hamiltonian can formally be written as
\begin{align}
	H(k_x)
	&=
	\sum_{n=1}^{N_n} \omega_n(k_x)\ket{k_x,n}\bra{k_x,n}
	\notag\\
	&+ \int d\alpha \, \epsilon_{\alpha}(k_x)\ket{k_x,\alpha}\bra{k_x,\alpha},
\end{align}
where $n$ labels the $N_n$ discrete subgap bands and $\alpha$ the continuum states. In our case with two subgap bands
we have $N_n = 2$, but we keep this number general, yet finite, for the following discussion.

The extraction of a 1D Hamiltonian requires two steps,
the rather easy projection on subgap energies to remove the continuum states, and the elimination of the $y$ coordinate.

In the following we keep $k_x$ as a fixed parameter and omit it from the notation for simplicity, without loss of generality.
The energy projection results in the Hamiltonian
\begin{equation}
	H'
	=
	\sum_{n=1}^{N_n} \omega_n \ket{n}\bra{n}.
\end{equation}
The states $\ket{n}$ span an $N_n$ dimensional subspace $\mathcal{H}'$ of the Hilbert space
$\mathcal{H}_{Ns} \otimes \mathcal{H}_y$, where $\mathcal{H}_{Ns}$ is the Nambu-spin space and
$\mathcal{H}_y$ is the space of square integrable functions of $y$.

We then seek a mapping $\Omega$ between operators on $\mathcal{H}'$ and operators on $\mathcal{H}_{Ns}$
such that $H_\text{eff}^\text{1D} = \Omega(H')$.
We impose the following two conditions such a mapping needs to fulfil:
\begin{enumerate}[(a)]
\item
The expectation values of any operator $A$ on $\mathcal{H}_{Ns}$, acting with the identity on $\mathcal{H}_y$,
 must remain invariant. This means we impose
\begin{equation} \label{eq:cond_a}
	\bra{n} A \ket{n'} = \Tr\{ \ket{n'}\bra{n} A \} = \Tr\{ \Omega(\ket{n'}\bra{n}) A\}.
\end{equation}
Notice that $A$ is kept outside the $\Omega$ mapping, which is not a physical requirement
but the choice of convenience mentioned above.
\item
For each orthogonal projector $\ket{n}\bra{n}$ the mapping produces again
an orthogonal projector, $\Omega(\ket{n}\bra{n}) = \ket{u_n}\bra{u_n}$,
with $\ket{u_n}$ in $\mathcal{H}_{Ns}$ such that $\bracket{u_n}{u_{n'}} = \delta_{n,n'}$.
\end{enumerate}
Condition (a) is the more stringent one, but condition (b) is the physical requirement as it
ensures that $H_\text{eff}^\text{1D}$ remains a Hamiltonian
on $\mathcal{H}_{Ns}$ with a spectral decomposition and the same spectrum. An immediate necessary
condition for (b) is that $N_n \le \dim(\mathcal{H}_{Ns})=4$.

To evaluate the consequences of condition (a) let us choose a set of states $\ket{\phi_p} \in \mathcal{H}_y$, for $p=1,\dots, N_p$, representing functions $\phi_p(y)$ such that
\begin{equation}
	\ket{n} = \sum_{p=1}^{N_p} \ket{v_n^p} \otimes \ket{\phi_p},
\end{equation}
with  $\ket{v_n^p}$ in $\mathcal{H}_{Ns}$.
We assume that $N_p$ is finite, and we see from Eqs.\ \eqref{eq:g_k_y}--\eqref{eq:p_sigma_pm} that the $\phi_p(y)$
indeed are expressed by the small set of functions $\exp(\pm i |y| k_F p_{\sigma,\pm})$. Through an orthogonalization procedure such
as the Gram-Schmidt method we can choose the $\ket{\phi_p}$ to be orthonormal, $\bracket{\phi_p}{\phi_{p'}} = \delta_{p,p'}$.
The normalization imposes furthermore that $\bracket{v_n^p}{v_n^p} = 1$ but otherwise there is no requirement for
orthogonality on the $\ket{v_n^p}$.
Equation \eqref{eq:cond_a} is then equal to
\begin{equation}
	\bra{n} A \ket{n'}
	= \sum_{p=1}^{N_p} \bra{v_n^p} A \ket{v_{n'}^p}
	= \Tr\Bigl\{ \sum_{p=1}^{N_p} \ket{v_{n'}^p}\bra{v_n^p} A \Bigr\}.
\end{equation}
This relation must hold for any $A$ and consequently
\begin{equation} \label{eq:Omega}
	\Omega(\ket{n'}\bra{n})
	= \sum_{p=1}^{N_p} \ket{v_{n'}^p} \bra{v_n^p}
	= \sum_{p=1}^{N_p} V_p \ket{n'} \bra{n} V_p^\dagger.
\end{equation}
The mapping $\Omega$ therefore takes the form of a Kraus decomposition \cite{Kraus1971,Choi1975}
with the Kraus operators $V_p = \openone_{Ns} \otimes \bra{\phi_p}$, where $\openone_{Ns}$ is the
identity on Nambu-spin space. Noting that $\sum_p V_p^\dagger V_p$ produces the identity on $\mathcal{H}'$
we find that $\Omega$ falls in the remit of Choi's theorem \cite{Choi1975}, which states that any
linear mapping from bounded operators acting on $\mathcal{H}'$ to operators acting on $\mathcal{H}_{Ns}$
that is completely positive and trace preserving is necessarily of the form of Eq.\ \eqref{eq:Omega}.

The minimum number $N_p$ of necessary Kraus operators is known as the Choi rank, but otherwise the
$V_p$ can be freely chosen as long as they fulfil Eq. \eqref{eq:Omega} and the identity condition on $\mathcal{H}'$.

We turn then to condition (b) and ask which choice of Kraus operators can guarantee the correct
mapping of projectors, which thus has to take the form
\begin{equation}
	\sum_{p=1}^{N_p} V_p \ket{n}\bra{n} V_p^\dagger
	= \sum_{p=1}^{N_p} \ket{v_n^p} \bra{v_n^p} = \ket{u_n}\bra{u_n}.
\end{equation}
Since $\dim(\mathcal{H}_{Ns})=4$ we can represent $\ket{v_n^p}$ as a $4 \times N_p$
matrix $\mathcal{V}_n$, and $\ket{u_n}$ as a length 4 column vector $\mathcal{U}_n$ such that
the latter equation becomes $\mathcal{V}_n \mathcal{V}_n^\dagger = \mathcal{U}_n \mathcal{U}_n^\dagger$.
This means $\mathcal{V}_n$ needs to be of rank 1, and therefore all its columns are directly linearly dependent.
In this case we have $\ket{v_n^p} = \lambda_n^p \ket{u_n}$ where the $\lambda_n^p$ are numbers such that
$\sum_{p=1}^{N_p} |\lambda_n^p|^2 = 1$. This, however, also imposes that
\begin{equation} \label{eq:separable}
	\ket{n} = \ket{u_n} \otimes \sum_{p=1}^{N_p} \lambda_n^p \ket{\phi_p}
	\equiv \ket{u_n} \otimes \ket{\psi_n}.
\end{equation}
This result shows that conditions (a) and (b) are only compatible if the states $\ket{n}$ are
separable in the sense of Eq.\ \eqref{eq:separable} in that for each $n$ the $y$ dependence is in a
single function $\psi_n(y) = \bracket{y}{\psi_n}$ multiplying the Nambu-spin states $\ket{u_n}$.
The $\ket{u_n}$ must be orthogonal but there is no orthogonality condition on the $\ket{\psi_n}$,
only normalization as $\bracket{\psi_n}{\psi_n} = 1$. Note that Eq.\ \eqref{eq:separable} does not imply that the Choi
rank is $N_p=1$ as the $\ket{\psi_n}$ can be different for different $n$.

For separable $\ket{n}$ the mapping $\Omega$ becomes then particularly simple and results in just
tracing out of the $y$ degrees of freedom,
\begin{multline}
	H_\text{eff}^\text{1D}
	= \Tr_y\{ H' \}
	= \int dy \, \bra{y} H' \ket{y}
\\
	= \sum_{n=1}^{N_n} \omega_n \ket{u_n} \bra{u_n} \int dy |\bracket{y}{\psi_n}|^2
	= \sum_{n=1}^{N_n} \omega_n \ket{u_n} \bra{u_n}.
\end{multline}
This result is remarkable in the sense that it confirms that for separable wave functions the elimination of the confining
degree of freedom by the intuitive simple integration is indeed the \emph{only} way that does not change the physics of the other degrees of freedom.
Separability is also encountered often for wave functions confined
by some potential such as created by a heterostructure, or of edge states in quantum Hall systems, topological insulators, or topological superconductors, in which the envelope does not depend on spin, and in which futher spatially dependent interactions that can hybridize the states are absent or negligible. In such a case it is straightforward to integrate out the spatial dependence and obtain an effective lower dimensional Hamiltonian for the bound states only.

On the other hand, if the states are not separable a Hamiltonian $H_\text{eff}^\text{1D}$ satisfying
both conditions (a) and (b) cannot be constructed. This indeed the general case
for the subgap states at the magnetic chain.
As mentioned before this

is seen from Eqs.\ \eqref{eq:g_k_y}--\eqref{eq:p_sigma_pm} through the amplitudes $\xi_\sigma$ and $\chi_\sigma$,
and their dependence on $p_{\sigma,\pm}$. For $|\omega|<0$ the latter satisfy $p_{\sigma,+}=p_{\sigma,-}^*$.
If we thus let $p_{\sigma,\pm} = p_{\sigma} \exp(\pm i \varphi)$ for $p_\sigma = |p_{\sigma,+}|$ and
$\varphi = \pm \mathrm{arg}(p_{\sigma,\pm})$, we see that
\begin{align}
	\xi_\sigma &= 2p_{\sigma}^{-1} \cos(y k_F p_{\sigma} \cos(\varphi) - \varphi) e^{- |y| k_F p_\sigma \sin(\varphi)},
\label{eq:xi_sigma_phi}
\\
	\chi_\sigma &= 2 p_{\sigma} \cos(y k_F p_{\sigma} \cos(\varphi) + \varphi) e^{- |y| k_F p_\sigma \sin(\varphi)}.
\label{eq:chi_sigma_phi}
\end{align}
As long as $p_{\sigma,\pm}$ is complex the division and multiplication by $p_{\sigma,\pm}$ adds opposite
phase offsets $\pm \varphi$ to the $y$ dependent oscillations of $\xi_\sigma$ and $\chi_\sigma$, so that the
$y$ dependence is not globally factorizable from the different terms of the wave function,
thus violating the separability of the wave function. The imaginary
part of $p_{\sigma,\pm}$ is furthermore required for the exponential confinement and exists whenever
$|\omega|<\Delta$.

To substantiate that indeed these factors of the Green's functions provide the relevant amplitudes
of the wave function let us note that we can write
\begin{equation}
	\bracket{y}{k_x,n} \bracket{k_x,n}{y'} = \oint_{C_{k_x}} \frac{d\omega}{2\pi i} G(\omega,k_x,y,y'),
\end{equation}
where $C_{k_x}$ is a positively oriented closed contour encircling only the isolated pole $\omega_n(k_x)$ of the Green's function.
Since at $|\omega|<\Delta$ the pole arises from the $T$ matrix we have
\begin{align}
	\bracket{y}{k_x,n} \bracket{k_x,n}{y'} = &g(\omega_n(k_x),k_x,y) \mathrm{Res} T(\omega_n(k_x),k_x)
\notag\\
	&\times
	g(\omega_n(k_x),k_x,-y'),
\end{align}
with $\mathrm{Res}T$ the residue of the $T$ matrix. Any $y$ dependence is thus due to $g(\omega_n(k_x),k_x,y)$ and any $y'$ dependence
to $g(\omega_n(k_x),k_x,-y')$. Hence the Green's functions $g$ directly define the $y$ dependence of the wave function, containing
the exponential envelopes and the oscillations. As they are
not separable in the sense above, the subgap states do not allow the reduction to an effective 1D Hamiltonian.

We should stress, however, that the lack of separability requires that the effect of the difference between the
$p_{\sigma,\pm}$ is notable, and situations can exist in which approximate separability and thus an approximately
valid 1D Hamiltonian can be obtained. Such a situation occurs when the exponential decay is fast compared
with the oscillation period, expressed by the condition $\mathrm{Im}p_{\sigma,+} \gg \mathrm{Re}p_{\sigma,+}$.
From Eq.\ \eqref{eq:p_sigma_pm} we see though that in the topologically most interesting limit of $\omega \to 0$
this condition does not hold. We then instead must consider the situation in which the phase shift $\varphi$ making
the oscillations of $\xi_\sigma$ and $\chi_\sigma$ distinct is negligible.
Since the characteristic range over which $y$ is evaluated is set by the decay length $1/k_F p_\sigma \sin(\varphi)$
we see from Eqs.\ \eqref{eq:xi_sigma_phi} and \eqref{eq:chi_sigma_phi} that the phase difference $\pm \varphi$ can be neglected when
$\cot(\varphi) \pm \varphi \approx \cot(\varphi)$, which is the case when $\cot(\varphi) \gg 1$. This represents thus
the limit $\mathrm{Im}p_{\sigma,+} \ll \mathrm{Re}p_{\sigma,+}$, which is precisely the limit in which the
long wavelength approximation (LWA) is applicable (see Part I). Full separability is then still not guaranteed as long
as $p_{\sigma,\pm}$ have different spin $\sigma$ dependence. But at the topologically most significant $k_x=0$
this spin dependence drops out and an approximate 1D Hamiltonian can be obtained by integrating out the $y$
dependence. This property confirms why this method of obtaining such a Hamiltonian produces valid results
in the LWA limit.

On the other hand, as discussed in depth in Part I, the range of applicability of the LWA becomes more and
more restricted for increasing $k_m$ and breaks down entirely at $k_m = k_F$, at which indeed
$\mathrm{Im}p_{\sigma,+} = \mathrm{Re}p_{\sigma,+}$ for $k_x=0$.
For the topological classification of the subgap states we therefore need a different approach which
we will describe next.


\subsection{Dimensional embedding}\label{sec:1D_in_2D}

Although it is not possibile to obtain an effective 1D Hamiltonian the wave functions remain 1D and we can expect that still some adjustment of the 1D topological classification schemes remains applicable. We thus aim to extract a 1D Hamiltonian solely for the purpose of the topological classification at the expense of removing any other physical significance. To this end it is useful to examine the analogy of how 1D topological invariants arise as weak 2D topological indices in particular directions.
For comparison we consider the example provided in Ref.\ \cite{Nagaosa2012} through a generalized model of a $p+ip$ superconductor on a 2D square lattice. Instead of performing a full 2D analysis, in this paper one of the momentum components $k_x$ or $k_y$ is treated as a fixed parameter and tuned to a time-reversal invariant point. In terms of the other momentum the Hamiltonian describes an effective 1D system, which in this case is equivalent to the Kitaev chain of a topological triplet superconductor. For the latter the topological classification is determined in the standard 1D way, and the obtained topological indices are identified with the weak topological 1D indices of the 2D system. The combination of the weak indices provides the characterization of the full 2D system. The effective 1D Hamiltonians do not necessarily have any direct physical significance but capture the topology at the significant time-reversal symmetric points. Since the system is translationally invariant these points are labelled by the momenta $k_x$ and $k_y$.

We are aiming for a similar extraction of an effective topological Hamiltonian. But due to the lack of translational symmetry along $y$ such a momentum space extraction of 1D Hamiltonians is not possible. To obtain the correct modification let us recall the role of time-reversal symmetric points. In a fermionic system with time-reversal symmetry each eigenstate has an orthogonal Kramers partner, its time reversed counterpart of opposite momentum and equal energy. At a time-reversal symmetric point the momenta of the Kramers partners coincide but their orthogonality prevents them from hybridizing and lifting the energy degeneracy. Only if more than one Kramers pair is present is a hybridization possible between states not belonging to the same pair, and only in the presence of an even number of Kramers pairs can the degeneracy be lifted entirely. The parity of the number of Kramers pairs is expressed through the $\mathbb{Z}_2$ index associated with the time-reversal symmetric point, and the impossibility to hybridize defines a topologically nontrivial state. Although most of the considered 1D topological systems involve some magnetic elements breaking time-reversal there is throughout either an emergent or an effective time-reversal symmetry \cite{Beck2021} for the relevant states so that the $\mathbb{Z}_2$ classification remains a valid standard tool. A similar choice, yet without any justification of the used topological Hamiltonian, was applied for a tight-binding model at $y=0$ in Ref.\ \cite{Sedlmayr2021}.

For the present case and in the limit of a large bulk gap $\Delta$ the wave functions are in the $y$ direction confined essentially to the magnetic chain position. The time invariant point is then given by $k_x=0$ and, through the confinement, by $y=0$. A classification through a topological Hamiltonian has to focus on this point. For a smaller $\Delta$ the wave functions widen around $y=0$ but any motion is still possible only in the $x$ direction. The relevant time-reversal points remain $k_x$ and $y$ dependent. We notice that the operation of time-reversal on the $y$ dependence of the Green's function is to transform the latter as $G(y,y') \to G(y',y)$, and time-reversal invariance requires thus $y'=y$. This includes the chain centre $y=y'=0$ which will provide the primary criterion for the topological classification. But it further allows the characterization at $y=y'\neq 0$. As discussed in Sec.\ \ref{sec:absence_1D_H} we must not integrate out the $y$ dependence, and instead below we will explore it further.


Consequently we define the $y$ dependent family of topological Hamiltonians through \cite{CarrollPhD2019}
\begin{align}\label{eqn:1DHam}
	H^\text{top}_\text{1D}(y) = - \left[G(\omega=0, k_x=0, y, y'=y)\right]^{-1},
\end{align}
where $\omega = 0$, $k_x = 0$ and $y = y'$ are chosen to fulfil the necessary symmetry conditions of
particle-hole symmetry at time-reversal invariant points in configuration space. The inverse is taken of
the $4 \times 4$ matrix $G(0,0,y,y)$.

The Hamiltonians $H^\text{top}_\text{1D}(y)$ represent a class of Hamiltonians obtained by slicing the 2D system
into effective 1D segments at a distance $y$ from the impurity chain. In this sense they are similar to the effective 1D Kitaev chain type Hamiltonians used for the determination of the weak 1D indices in the bulk system, with $y$ replacing the use of a momentum as parameter. But the $y$ parameters are not limited to special values as time-reversal symmetry is built in through $y'=y$ in the Green's function, and $y$ is tunable through all values. We will show that these Hamiltonians correctly produce the topological behaviour of the subgap states in the vicinity of $y=0$, reproducing the topological phase diagram of the pure 1D chain when taking into account the renormalized critical coupling strengths.
The correctness of the Green's function at all wavelengths emphasized in Part I is of crucial importance here for the validity of the phase diagram, as could already be deduced from its significance on the non-separability of the $y$ dependent wave function discussed in Sec.\ \ref{sec:absence_1D_H}.

As the subgap bands are exponentially localized at the chain, the topology at large $y$ must become trivial.
Since the topological indices are integers the passage to a trivial topology has to be abrupt and there must exist an effective
boundary between the region near the chain and the rest of the superconductor. Through $H^\text{top}_\text{1D}(y)$ we can
capture this behaviour, but we should emphasize that $H^\text{top}_\text{1D}(y)$ is only to be taken as an archetypical
representative of $y$ dependent topological Hamiltonians. The pure topological and not physical interpretation is furthermore underlined by noting that in addition to the symmetry considerations the classification depends on the change of the sign of eigenvalues about the Fermi level and not necessarily on the eigenvalues passing through the Fermi level \cite{Gurarie2011,Wang2012a,Wang2012b,Wang2012c,Wang2013,Budich2013,Weststrom2016}.
Since the states do not change, the transition to the trivial phase with increasing $y$ indeed cannot rely on Fermi level crossings and, as further investigated below, is instead bound to divergences in the spectrum of $H^\text{top}_\text{1D}(y)$ due to zeros in the defining Green's function,
which themselves are the expressions of nodes in the subgap wave functions.

Before continuing we should mention that alternative classification methods for spatially inhomogeneous systems
were put forward several years ago in the form of local Chern markers \cite{Bianco2011}, the Bott index \cite{Hastings2011},
and non-commutative Chern numbers or Chern number densities \cite{Prodan2010,Prodan2011,Mascot2019a,Mascot2019b}. Such
quantities allow spatial variations in the topological classification.
These approaches replace the derivatives in momentum space for the usual Chern numbers by traces over local coordinates
in real space together with projections onto occupied states. We found though that for our current purpose the
method we propose is more readily accessible and provides the correct topological classification.

\subsection{Topological classification near the chain}

\begin{figure}
	\centering
	\includegraphics[width=\columnwidth]{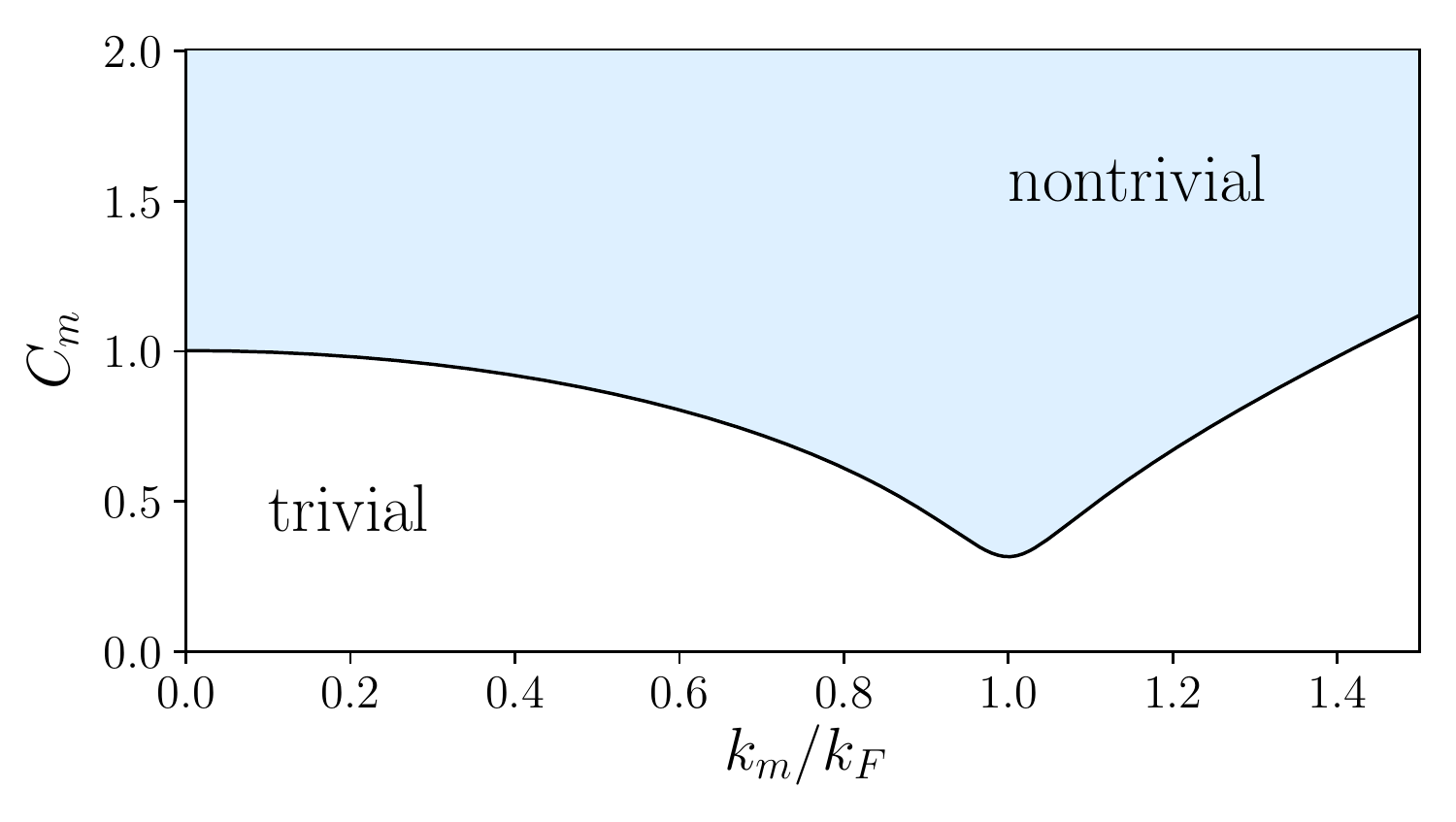}
	\caption{\label{fig:2D_topol_phasediagram}%
	Topological phase diagram obtained from the topological Hamiltonians $H^\text{top}_\text{1D}(y=0)$ as a function of spiral wave
	number $k_m$ and magnetic scattering strength $C_m$, for $\tilde{\Delta}=0.1$.
	This diagram corresponds to the phase diagram of the pure 1D model of Fig.\ \ref{fig:1D_phase_diagram}, with the same colour coding,
	upon the discussed dimensional renormalization, with values $C_m^\star$ [Eq.\ \eqref{eqn:kx0gapclosure}] marking the transition
	by the solid line, instead of the values $\check{C}_m^\star$ [Eq.\ \eqref{eqn:1D_phase_boundary}].
	Notably the transition at $k_m=k_F$ is now at the larger $C_m = \tilde{\Delta}^{1/2}$ instead of $C_m \propto \tilde{\Delta}$.
	}
\end{figure}

\begin{figure}
	\centering
	\includegraphics[width=\columnwidth]{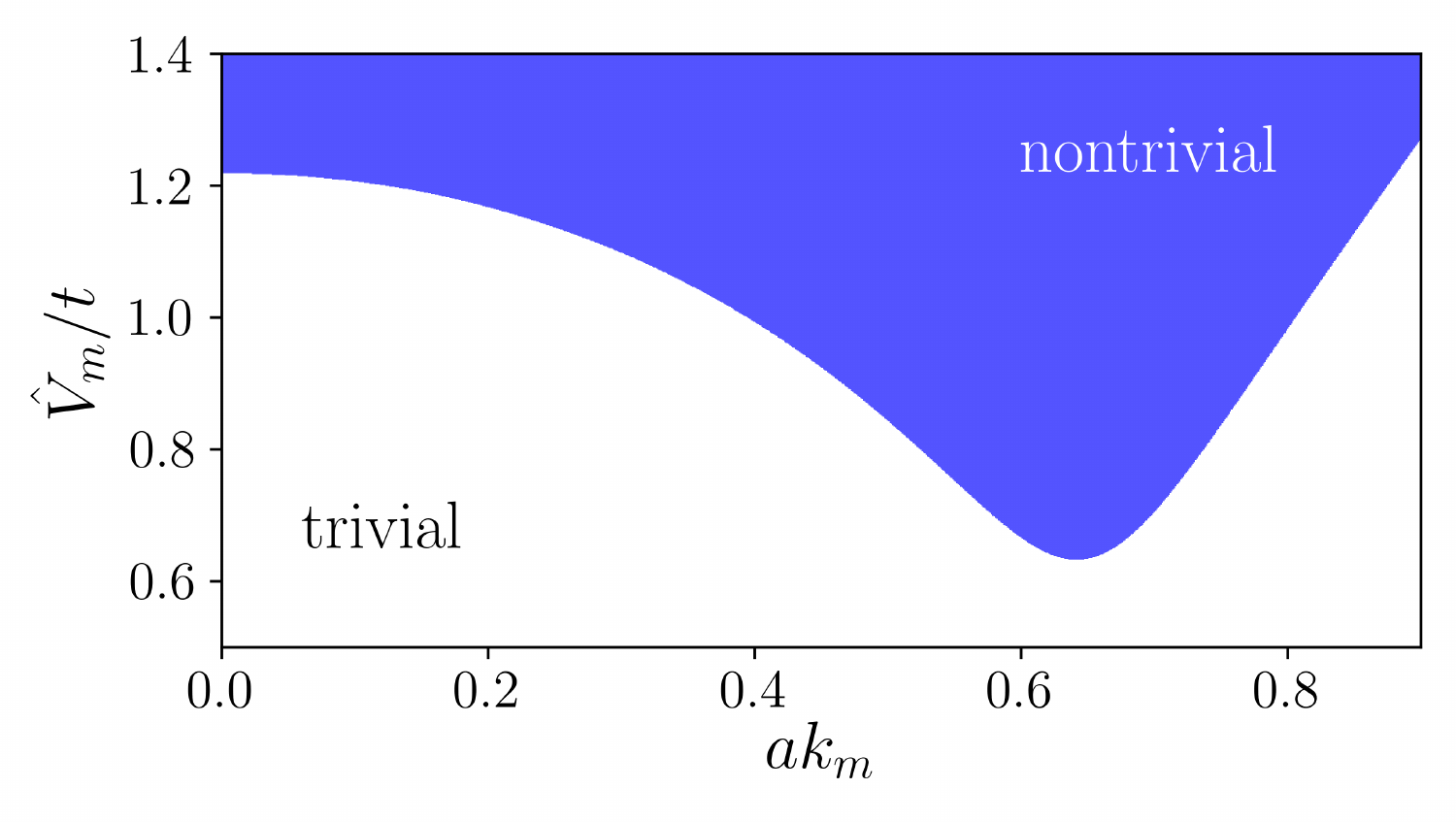}
	\caption{\label{fig:topol_phasediagram_num}%
	Topological phase diagram obtained from the self-consistent numerical solution of the matching tight-binding model
	described in Appendix \ref{app:numerics}, as a function of spiral wave number $k_m$ and magnetic scattering strength
	$\hat{V}_m$. Scales are given in units of the hopping integral $t$ and the lattice constant $a$.
	The pairing interaction and chemical potential are chosen to produce $\Delta \approx 0.1 t$ and $k_F a \approx 0.65$.
	All the features of the analytic model are perfectly reproduced, only the numerical values of $\hat{V}_m$ are not directly
	comparable with $C_m$ because of the involved different density of states and effective mass.
	}
\end{figure}

\begin{figure}
	\centering
	\includegraphics[width=\columnwidth]{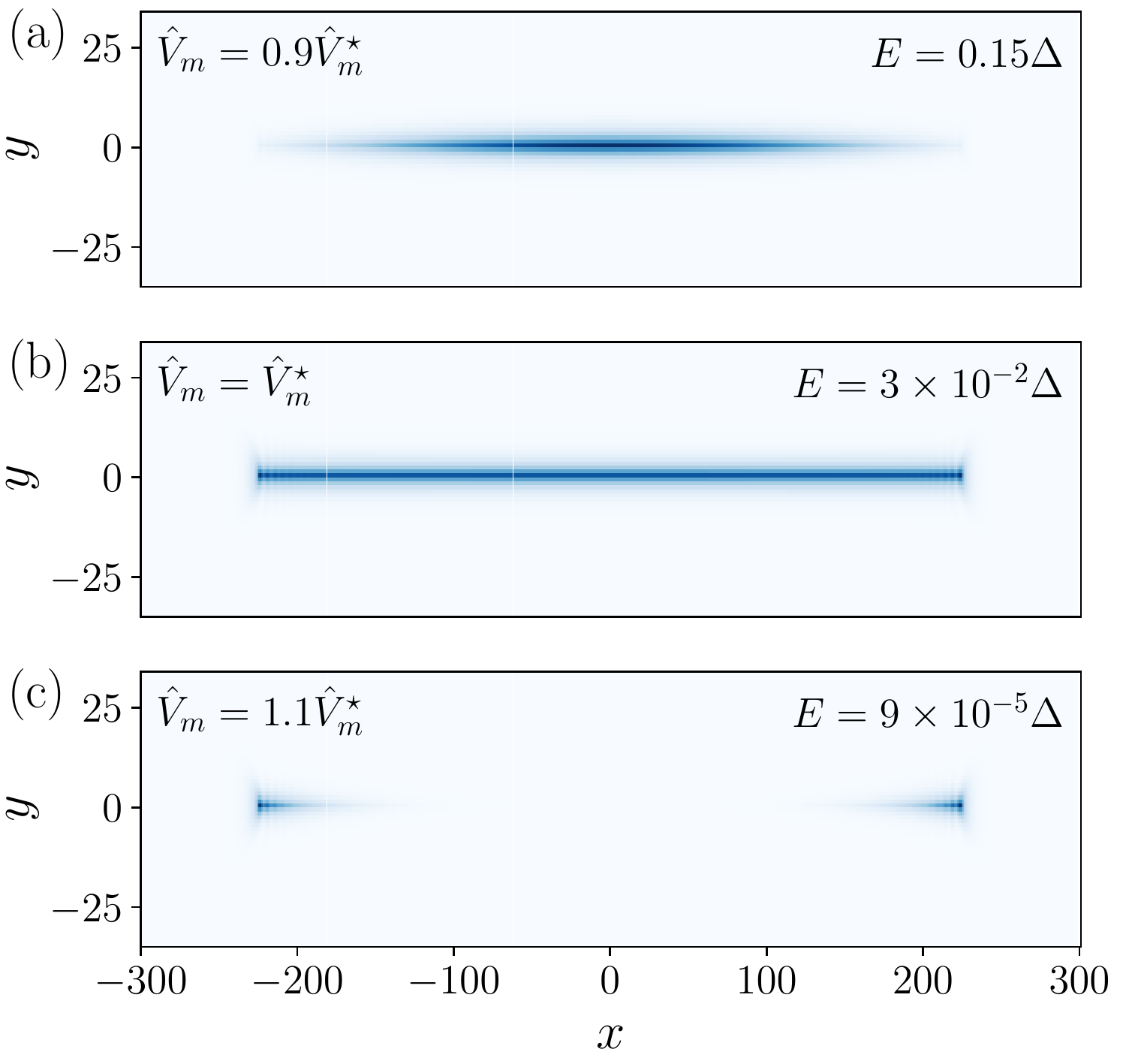}
	\caption{\label{fig:lowest_eigval}%
		Real space map of the absolute square of the wave function for the smallest
		eigenvalues of a real space system of $600 \times 70$ sites with a spiral magnetic chain extending
		between sites $x=-224$ to $x=225$ at $y=0$, with spiral wave vector $k_m = k_F$. Darker pixels show a larger
		amplitude, the eigenenergies are $\pm E$ as shown in the panels. The shown amplitudes are summed over spin and particle-hole
		components, and for better visualization of the end states we have summed furthermore over both the $+E$ and $-E$ amplitudes.
		The magnetic impurity
		potentials $\hat{V}_m$ are chosen to lie (a) below, (b) at, and (c) above the gap closing strength
		$\hat{V}_m = \hat{V}_m^\star = ( 4 t \Delta)^{1/2}$, with $t$ the hopping integral and $\Delta=0.1 t$ the gap function.
		Panel (c) demonstrates the topological nature of the transition through the appearance of the Majorana end states
		with energy $E \approx 0$ within the accuracy of the remaining finite size wave function overlap.
	}
\end{figure}

Since $H^\text{top}_\text{1D}(y)$ are matrices in Nambu-spin space their topological classification is most easily done
through the Pfaffians at time-reversal symmetric points, which for the continuum model is reduced to the
behaviour at $k_x=0$ in the $k_m$ shifted basis.
The relevant topological index is then as in the 1D case above determined by the sign of $\pf[ UH^\text{top}_\text{1D}(y)]$ \cite{Stanescu2011},
where the matrix $U=\sigma_x \tau_x$ again transforms the Hamiltonian to a skew symmetric matrix.
In Fig.\ \ref{fig:2D_topol_phasediagram} we plot the resulting topological phase diagram for the topological
Hamiltonian at the position $y=0$ of the impurity chain
as a function of spiral winding $k_m$ and dimensionless magnetic interaction strength $C_m$ [see Eq.\ \eqref{eqn:C_m}].
The shaded areas are the topologically nontrivial range.
In comparison with Fig.\ \ref{fig:1D_phase_diagram} we see that the results perfectly reflect
the phase diagram of the pure 1D system under the aforementioned dimensional renormalization.
The phase transition occurs when the subgap bands touch at the Fermi level at $k_x=0$. This is exactly at the critical interaction strength $C_m^\star$ given in Eq.\ \eqref{eqn:kx0gapclosure} which replaces the $\check{C}_m^\star$ of the pure 1D system of Eq.\ \eqref{eqn:1D_phase_boundary}.
As there is no other gap closing at $k_x=0$ and for the continuum model there is no finite momentum at the edge of the Brillouin zone there is no mechanism for a phase transition at any other interaction strength.

To corroborate the validity of these results by an independent method we compare them with the numerical solution of the
tight-binding model that has already provided excellent quantitative verification in Part I. We perform two validations,
the first by comparing the matching topological invariants, and the second by demonstrating the appearance of zero modes
localized at the edges of a finite chain.

For the first verification we also use the Pfaffians of the topological Hamiltonians for which we compute the Green's functions through their Lehmann representation from the eigenvalues and eigenvectors of the full 2D Hamiltonian. Appendix \ref{app:numerics} contains a further
description of the numerical evaluation.
The numerical results are shown in Fig.\ \ref{fig:topol_phasediagram_num}, in which we again plot the diagram as function of $k_m$ and the magnetic scattering strength which we denote for the tight-binding model by $\hat{V}_m$.
The agreement is excellent as the phases and the shape of the phase transition line are perfectly matched. We should only note that
the numerical values of $\hat{V}_m$ for the transition are not the same because the densities of state of the two models are different.
We remark furthermore that for the tight-binding model we have only considered $k_x=0$ and not its second time-reversal symmetric point $k_x=\pi/a$ at the edge of the Brillouin zone, as the latter is absent in the continuum model.
We thus exclude in the tight-binding model the possibility to leave the topological phase at large $C_m$ due to a gap closing at $k_x=\pi/a$.

The second verification of the validity of the topological classification through $H^\text{top}_\text{1D}(y=0)$
is shown in Fig. \ref{fig:lowest_eigval}. In this figure we display for spiral wave vector $k_m=k_F$ how the wave functions of the
eigenvalues $\pm E$
closest to the Fermi level change from an extended 1D state to localized end states when $\hat{V}_m$
changes across the gap closing interaction strength $\hat{V}_m^\star$ corresponding to $C_m^\star$ in the continuum model.
For better visualization we plot the sum of the amplitudes of the two wave functions for $\pm E$. Due to particle-hole
symmetry the amplitudes are the same for the extended states, and for the localized end states we assure in this way that
the states at both ends are visible. We verify furthermore that the values of $E$ (shown as labels in the figure)
decrease to $E=0$ within the numerical accuracy. Only these states are localized and we verified that the other eigenstates remain extended.
These end states are thus indeed the particle-hole symmetric Majorana bound states expected from a transition to
the topologically nontrivial phase.
Through these verifications we can thus confirm that the topological Hamiltonian $H^\text{top}_\text{1D}(y=0)$
indeed produces the correct topological classification.

\subsection{Topology at $y \neq 0$}

\begin{figure*}
	\centering
	\includegraphics[width=\textwidth]{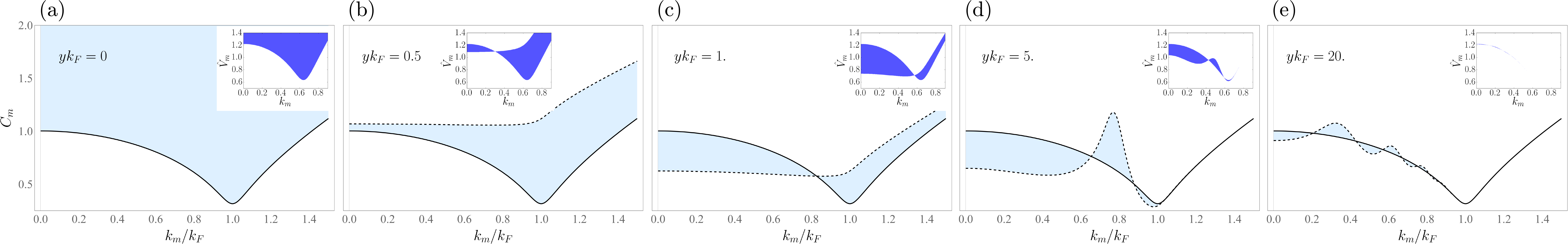}
	\caption{\label{fig:topol_phasediagram}%
	Topological phase diagrams obtained from the topological Hamiltonians $H^\text{top}_\text{1D}(y)$ as a function of spiral wave
	number $k_m$ and magnetic scattering strength $C_m$, for various $y$ and $\tilde{\Delta}=0.1$.
	(a) is identical to Fig.\ \ref{fig:2D_topol_phasediagram} and displays the principal phase diagram at $y=0$.
	The solid line shows the primary transition at $C_m^\star$ [Eq.\ \eqref{eqn:kx0gapclosure}] between the topologically trivial (white)
	and nontrivial (blue) regions.
	Panels (b)--(e) show with the dashed line the appearance at $y\neq 0$ of the second transition at
	strength $C_m^{\star\star}$ [Eq.\ \eqref{eqn:zerolocations}], determined by the zeros of the Green's function. At large $y$
	the region spanned between both lines shrinks to zero such that the system becomes trivial throughout. At intermediate distances
	oscillations of the dashed line about the solid line show that at the same interaction strength a region can change topology
	several times with $y$, and some trivial regions at $y=0$ can become nontrivial at some nonzero $y$.
	The insets display the corresponding diagrams for the numerical solution of the tight-binding model, and
	show a remarkable correspondence with the continuum model.
	Differences appear only in the magnitude of regions or are due to limitations of the discrete $y$ values on the lattice
	as in (b) where there is no lattice site close enough to the interface to directly match $yk_F=0.5$.
	The inset of (a) reproduces Fig.\ \ref{fig:topol_phasediagram_num}.}
\end{figure*}

\begin{figure*}
	\centering
	\includegraphics[width=\textwidth]{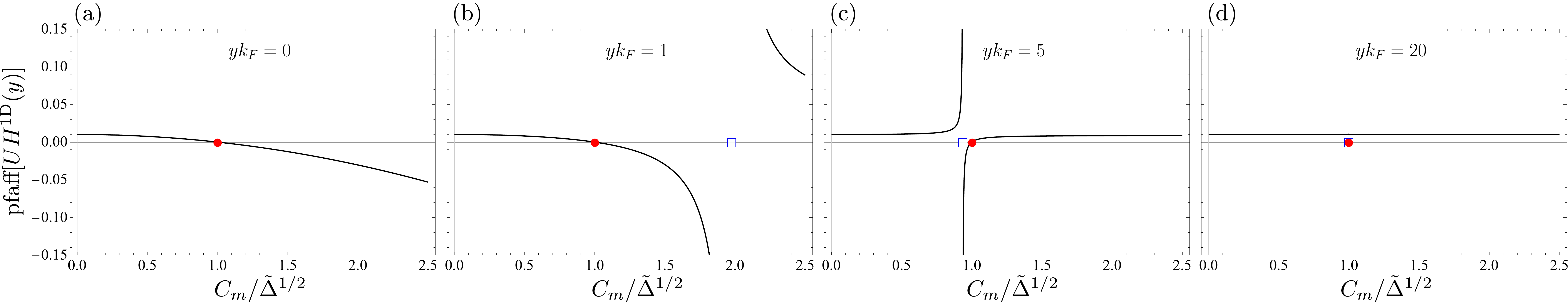}
	\caption{\label{fig:topol_index_range_ybar}%
	Pfaffian of the 1D Hamiltonian as a function of magnetic interaction strength $C_m$ for increasing distances $y$ similar to
	Fig.\ \ref{fig:topol_phasediagram}, but for fixed spiral wave number $k_m=k_F$. The interaction strengths are
	normalized to the critical $C_m = C_m ^{\star} = \tilde{\Delta}^{1/2}$ at which the gap closes at $k_x=0$.
	The Pfaffian changes its sign at both the zero at $C_m=C_m^\star$ (indicated by the red circle) and the
	pole at $C_m=C_m^{\star\star}$ (blue square) of $H^\text{top}_\text{1D}(y)$.
	The circle corresponds to the cut through the solid line and the square to the cut through the
	dashed line in Fig.\ \ref{fig:topol_phasediagram} at $k_m=k_F$.
	While $C_m^\star$ is independent of $y$, the value of $C_m^{\star\star}$ strongly varies with increasing $y$.
	At the large $y$ in (d)
	the overlap of zero and pole is well seen and the zero eliminates the divergence such that the curve
	is continuous throughout. This indicates
	the absence of any topological transition at large distances even at $C_m$ values at which panels (a)--(c) show the existence of
	a topologically nontrivial phase nearer the impurity chain.
	}
\end{figure*}

With the physical significance of the $y=0$ Hamiltonian verified, we inspect the further $y$ dependence.
Since the $H^\text{top}_\text{1D}(y)$ are a choice this analysis is principally only qualitative.
Nevertheless we find that the properties underlying the transition from the topology near the chain to the
trivial topology in the bulk are governed by physical and plausible mechanisms. For this reason we provide a
detailed analysis of the $y$ dependence, in particular as it reveals an interesting picture of the
extension of the topological regions into space. Furthermore, as we show below a leading role will be played by
the zeros of the Green's function (meaning $\det G = 0$ here) which is otherwise found only for interacting systems \cite{Gurarie2011,Volovik}.
Thus the family of 1D Hamiltonians $H^\text{top}_\text{1D}(y)$ can also be viewed as a simulator
of features that otherwise occur only in strongly correlated systems. Here we exhibit these features through
the means of $H^\text{top}_\text{1D}(y)$ but it could similarly be achieved by directly analyzing
$G(\omega,k_x,y,y)$ as a class of 1D Green's functions with an effective strong correlation physics whose
interaction strength is controlled by $y$.

We display the topological classification as a function of $y$ in Fig.\ \ref{fig:topol_phasediagram},
with $y=0$ in Fig.\ \ref{fig:topol_phasediagram}(a) repeating Fig.\ \ref{fig:2D_topol_phasediagram} for completeness, and with
increasing values of $y>0$ in Figs.\ \ref{fig:topol_phasediagram}(b)--(e). The insets show corresponding data from the numerical
solution of the tight-binding model, repeating Fig.\ \ref{fig:topol_phasediagram_num} in the inset of
panel Fig.\ \ref{fig:topol_phasediagram}(a).
At large values of $y$ the subgap states are all exponentially suppressed and we expect that $H^\text{top}_\text{1D}(y)$
exhibits only a topologically trivial phase. This is confirmed by Fig.\ \ref{fig:topol_phasediagram}(e)
which shows that the topological nontrivial region collapses far from the impurity chain.
It is interesting to analyze how this collapse occurs, and we observe in Figs.\ \ref{fig:topol_phasediagram}(b)--(e) that it is indeed far from
being simple. Most significant is in Fig.\ \ref{fig:topol_phasediagram}(b) the appearance of a second transition line at which for increasing $C_m$ the system
becomes again trivial. To understand this behaviour we should notice that the phase diagram of Fig.\ \ref{fig:topol_phasediagram}(a) results from the usual
crossing of the Fermi level of an eigenvalue of the Hamiltonian.

In terms of the Green's function a pole then crosses
the Fermi level, which coincides with the pole of the $T$ matrix. Since this pole is set by the interaction it is the same
for all $y$. This is shown by the solid line in all panels in Fig.\ \ref{fig:topol_phasediagram}.
The only way the sign of the Pfaffian can then change is when a zero of the Green's function instead of a pole
crosses the Fermi level, and the zeros of the Green's functions then mark the transitions to the trivial region at large $y$.
In Fig.\ \ref{fig:topol_phasediagram} we have marked the crossing of a zero of the Green's function by a dashed line
to distinguish it from the $y$ independent crossing of the pole shown by the solid line.
As $y$ increases the poles and zeros increasingly coincide, causing the topologically nontrivial region eventually to vanish.

To substantiate these statements let us look first at the condition $\pf[U H^\text{top}_\text{1D}(y)] = 0$. Since $\det(A) = \pf^2(A)$ for any
skew symmetric matrix $A$ this condition is indeed set by the divergence of $\det[G(0,0,y,y)]$. Such a divergence occurs through
the divergence of $\det[T(\omega=0,k_x=0)]$, which is precisely the condition for the existence of a subgap state at
frequency $\omega=0$ and momentum $k_x=0$ used in Part I for the characterization of the subgap spectrum.
Since $\omega=0$ this is the same condition as the gap closure condition at $k_x=0$, for which we have determined the critical
interaction strength $C_m^\star$ in Eq.\ \eqref{eqn:kx0gapclosure}.
Thus, very close to the interface, the phase transition is governed entirely by the poles of the Green's function.

As $y$ moves away from the interface the amplitude of the $T$ matrix
term in the Green's function at $\omega=0$ decays exponentially and $H^\text{top}_\text{1D}(y)\rightarrow - [g(0,0,0)]^{-1}$ which is topologically trivial.
Since the denominators of $G$ are $y$ independent the necessary change of sign of the Pfaffian of $H^\text{top}_\text{1D}(y)$ can no longer come from the crossing of a pole of $\det[G(0,0,y,y)]$. Instead it has to appear from a pole of $H^\text{top}_\text{1D}(y)$ itself, when one of the eigenvalues diverges, for instance, to $+\infty$ and reappears at $-\infty$. Since $H^\text{top}_\text{1D}(y)$ is given by the inverse Green's function, the location of this pole corresponds to a zero of $\det[G(0,0,y,y)]$.
In Fig. \ref{fig:topol_index_range_ybar} we visualize this effect by plotting the value of the
Pfaffian against magnetic interaction strength $C_m$ for a range of $y$ for $k_m = k_F$. The position of the pole of the Green's function is shown by the circle and the position of the zero of the Green's function (at $y \neq 0$) by the square. For increasing $y$ the pole and zero converge until they overlap and the system remains topologically trivial for all interactions strengths.

\begin{figure}
	\centering
	\includegraphics[width = \columnwidth]{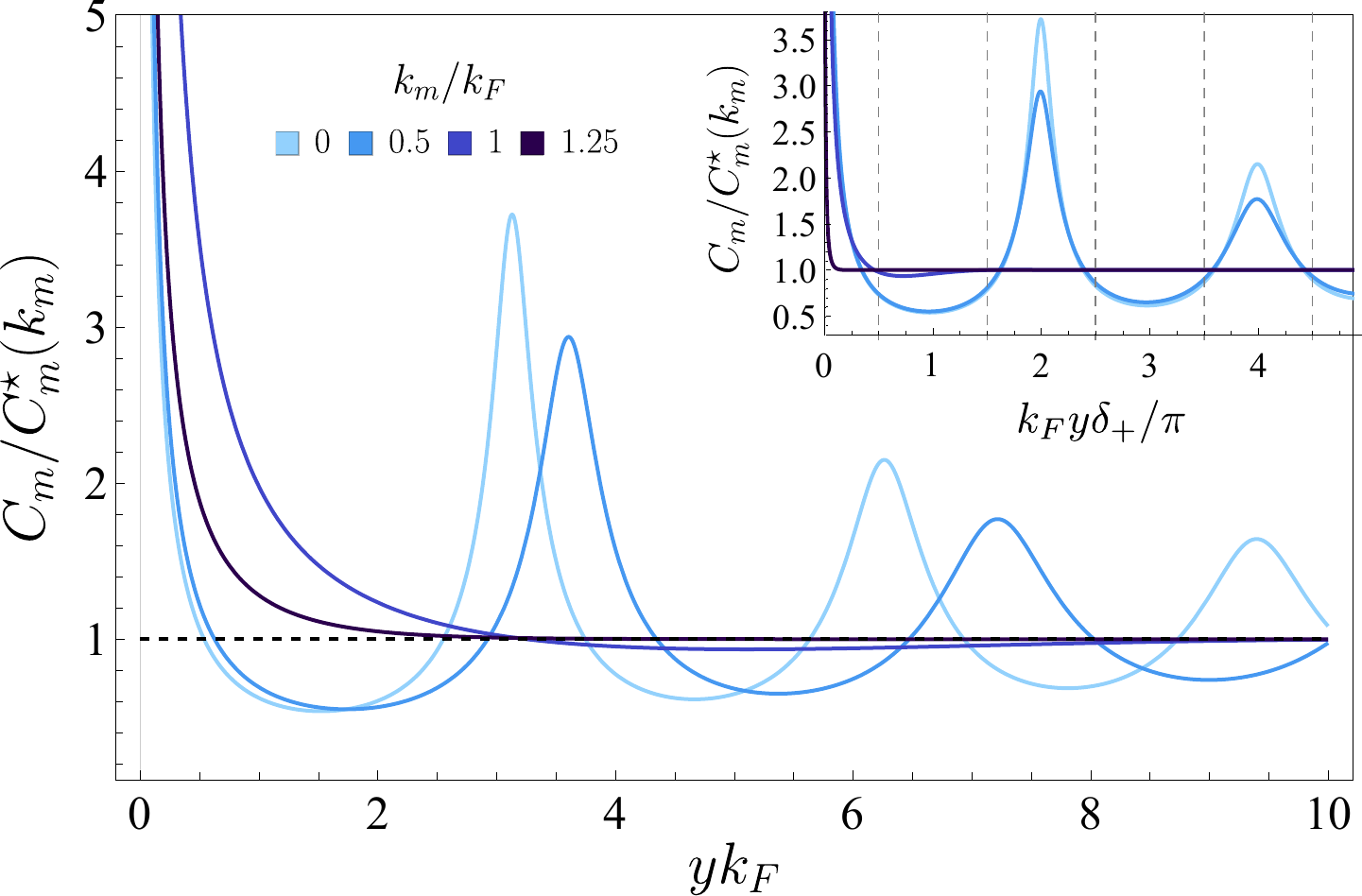}
	\caption{\label{fig:topol_polelocation_kmkf}Plot of the interaction strengths for the zeros $C_m^{\star\star}(y)$ (coloured, thick curve)
	and poles $C_m^\star$ (black, dashed curve, $y$ independent) of the Green's function as a function of $y$ for a range of spiral wave vectors $k_m$.
	The $C_m$ axis is normalized to $C_m^\star$ which depends on $k_m$. The inset displays the same functions with $y$ normalized to
	the dimensionless oscillating scale $y k_F \delta_+/\pi$ in Eq.\ \eqref{eqn:zerolocations}. The plots illustrate the generality
	of the topological strips and the possibility to enter a non-topological phase remotely from the impurity chain (within enclosed regions
	 between the coloured, thick curves and black, dashed curve).}
\end{figure}

The condition $\det\left[G(0,0,y,y)\right] = 0$ actually admits an exact solution for the location of this pole in the Pfaffian.
From the exact, full Greens function defined in Eq.\ \eqref{eq:G} we obtain
\begin{align}\label{eqn:zerolocations}
	C_m^{\star\star}
	= \frac{C_m^\star}%
	       {\sqrt{1 + e^{-2|y| k_F \delta_-} - 2e^{-|y| k_F \delta_-}\cos\left(|y| k_F\delta_+\right)}},
\end{align}
where $\delta_\pm = \sqrt{2} [(C_m^\star)^2 \pm (1- k_m^2/k_F^2)]^{1/2}$ and $C_m^{\star}$ is the magnetic interaction strength at which the Green's function admits a pole, as defined in Eq.\ \eqref{eqn:kx0gapclosure}.
The value $C_m^{\star\star}$ completely determines the additional, dashed phase boundary in
Fig. \ref{fig:topol_phasediagram} and is marked by the square in Fig. \ref{fig:topol_index_range_ybar}.
In Fig. \ref{fig:topol_polelocation_kmkf} we show $C_m^{\star\star}$
in comparison with $C_m^\star$ as a function of $y$ for a selection of spiral wave vectors $k_m$.

Equation \eqref{eqn:zerolocations} shows that the topological phase diagram is governed by two dimensionless parameters. One set by
$y k_F \delta_-$ providing how $C_m^{\star\star}$ approaches $C_m^\star$ away from the interface as a function of $k_m$ and $y$,
and one set by $y k_F \delta_+/\pi$ describing the oscillations of $C_m^{\star\star}$ about $C_m^\star$. These length
scales arise from the natural scales of the Green's function given by Eq.\ \eqref{eq:g_k_y}.
Indeed we have $\delta_+ = \mathrm{Re}(p_{\sigma,\pm})$ and $\delta_- = \pm \mathrm{Im}(p_{\sigma,\pm})$,
where $p_{\sigma,\pm}$ is taken at $\omega=0$ and $k_x=0$ at which it is independent of $\sigma$ and $p_{\sigma,+}=p_{\sigma,-}^*$.
Therefore $\delta_+$ sets naturally the oscillatory behaviour in $C_m^{\star\star}$ and $\delta_-$ the exponential convergence at longer distances.

The oscillations of $C_m^{\star\star}$ about $C_m^\star$ lead to the interesting consequence observed in
Figs. \ref{fig:topol_phasediagram} and \ref{fig:topol_index_range_ybar} that when moving away
from the impurity chain the topological Hamiltonian $H^\text{top}_\text{1D}(y)$ changes its topological classification several times before
settling in the topologically trivial phase.
This means that there are strips near the chain that can
be considered as alternatively trivial and nontrivial, with a width of the strips set by half of the
oscillation scale $\Delta y \sim  \pi/k_F \delta_+$. The universality of the latter scale is shown
by the inset in Fig. \ref{fig:topol_polelocation_kmkf}.
We notice in particular that in Fig. \ref{fig:topol_phasediagram} (c), at $k_m \lesssim 0.8 k_F$,
entrance to the topological phase is triggered by a zero of the Green's function rather than a pole. This highlights the
fact that it is possible for strips at particular $y \neq 0$ to become nontrivial \emph{before} the interface at $y=0$
itself does as $C_m$ is tuned and without any requirements at all on subgap states.
This can be clearly seen in Fig.\ \ref{fig:topol_polelocation_kmkf} where there are large regions of space where
$C_m^{\star\star}<C_m^\star$ and which are thus nontrivial at only a fraction of the magnetic interaction strength
required at the interface. Additionally, there can be multiple $k_m$ points [for example around
Fig. \ref{fig:topol_phasediagram}(d) for $k_m \approx 0.65 k_F$ and $ \approx 0.87k_F$] where the
pole and zero coincide and hence the system is topologically trivial for any magnetic interaction strength $C_m$.

We should recall here that the topological Hamiltonians $H^\text{top}_\text{1D}(y)$ are only representative for the topological aspects
and do not allow a one-to-one matching with physical properties. Nevertheless they incorporate the natural scales and properties of the system as they are built from the physical Green's function, and as a function of $y$ they have a clear prediction of
alternating strips of topologically trivial and nontrivial regions of widths set by the natural scales of the system.
Taken as real objects there would be interfaces between strips of different topological classification and thus suggest the
existence of interface states at these interfaces. Since the interfaces are very close together these interface states
all overlap and produce a single wave function with spatial modulation corresponding to the strip widths that is
captured by the Green's function. Similar oscillating patterns appear in many other systems from scattering at
any interface or impurity in the form of Friedel
type oscillations. Some examples of oscillating densities and currents in superconductors are found in
Refs. \cite{Matsumoto1999,Wang2004,Horovitz2003,Braunecker2005,Kraus2008,Lauke2018}.
Although speculative it may thus be interesting to see if there could indeed be an interpretation of such
oscillating patterns that are found through conventional calculations in terms of the concept of
patterns of topologically distinct regions. Such a study is beyond the scope of the present paper.

On the other hand, the Green's function is a physical object that is principally measurable, allowing thus a direct
determination of the topological Hamiltonians. The spatial dependence of the subgap states near the magnetic chain
can then be used to continuously tune the Hamiltonians and their topology. Each Hamiltonian is then taken as a real
object that is simulated by the underlying superconducting system, and the principal topological properties are
determined by the zeros of the Green's function as a function of $y$. As the Green's function at fixed $y=y'$ is a
slice out of a higher dimensional system, it is renormalized by the nontrivial higher dimensional structure and thus
can incorporate structural changes that in a bulk system would require strong interactions, notably the appearance
of its zeros. Through such an interpretation the
discussion of the topological properties given above becomes a reality within the simulated model Hamiltonians.


\section{Conclusions}\label{sec:conclusions}

In this paper we investigated the topological properties of the subgap states appearing
in a superconductor through scattering on a chain of densely packed magnetic impurities
for ferromagnetic or spiral magnetizations. We demonstrated that it is necessary to go
beyond a straightforward topological classification attempt. To provide such a classification
the precise form of the Green's function as derived in Part I of this work becomes fundamentally
important as it allows one to set up a correct classification method that remains valid for
all scattering strengths $V_m$ (or the dimensionless $C_m$) and all magnetic spiral wave
numbers $k_m$.

We showed how the Green's function provides a precise prescription of the gap closures
at $k_x=0$ and we set up a family of topological Hamiltonians $H^\text{top}_\text{1D}(y)$ that captures at the position $y=0$
of the impurity chain the associated topological phase transitions at any $k_m$.
Through this approach we circumvented the difficulties we showed to arise from the attempt to extract
an effective physical Hamiltonian for the confined subgap states by conventional elimination
of the $y$ degree of freedom. It gave us the additional
benefit of obtaining a qualitative prescription of how a topologically nontrivial physics near
the chain transitions to the topologically trivial regions far from the chain, $y \to \infty$, where subgap states
are absent. This transition is necessarily driven by the zeros of the Green's
function which at large distances align with the singularities and in this way neutralize
any possible topological phase transition. The oscillations created by the
$y$ dependence of the Green's function therefore cause a behaviour mimicking the reduction
and vanishing of density of states of strongly correlated bulk systems that can also
produce a topological phase transition.
The $y$ dependence simulates such a behaviour and the analysis that is provided shows that it indeed
has to appear in systems of topologically nontrivial states that are confined in some
topologically trivial background to guarantee that the bulk topological phase is recovered
at large distances.

It should be emphasized though that in this case the zeros in the Green's function are not a consequence of the spectral function or wave
functions becoming zero. One can plot spectral functions through the transition and observe no obvious,
sharp change, in contrast to the case of poles of the Green's function where there is a discontinuity. Instead, the zeros are due
to a loss of linear dependence in the Green's function caused by competition between the magnetic interaction strength and the
background superconductor. This results in an emergent symmetry between states, expressed by the alignment of
a zero with a pole with increasing $y$, which can be compared to transitions governed by poles where states
move in frequency space and, by careful tuning, can coincide with high symmetry points in configuration space.
This property thus assures the fitness of these Hamiltonians for the spatially dependent topological classification.

Interestingly the spatial oscillations of the subgap wave functions can lead to the appearance of multiple
strips of different topological index in the vicinity of the chain.
This may be compared with layers of different materials, but due to the constructed nature of the topological Hamiltonians
any physical implications would remain speculative. In addition these layers are very
narrow, below the superconducting coherence length and Fermi wavelength, so that any features that could arise
from interfacing different materials would be washed out broadly through many layers.
Yet there are situations in which the topology near the impurity chain is trivial and a nontrivial strip
appears only at a distance. This raises the general question as to whether it could be possible to design spatial
patterns of regions with different topological properties by interference of such wave functions arising from an astute
placement of magnetic scatterers.

\acknowledgments
We thank T. Cren, R. Queiroz, T. Ojanen, C. Hooley and P. Simon for stimulating discussions, and A. V. Balatsky for discussions during
the early stage of this work.
CJFC acknowledges studentship funding from EPSRC under Grant No. EP/M506631/1.
The work presented in this paper is theoretical. No data
were produced, and supporting research data are not required.


\appendix

\section{Subgap bands from self-consistent numerics}\label{app:numerics}

We employ the tight-binding model introduced in Part I for comparison with the analytical model and
validation of the results. This model is defined through the Hamiltonian
\begin{align}
H
&=
- \sum_{\langle i,j \rangle, \sigma} t c_{i,\sigma}^\dagger c_{j,\sigma}
- \sum_{i, \sigma} \mu c_{i,\sigma}^\dagger c_{i,\sigma}
\notag\\
&+ \sum_{i} \left[ \Delta_{i} c_{i,\downarrow} c_{i,\uparrow} + \text{h.c.} \right].
\end{align}
The indices $i,j$ run over the sites of a 2D square lattice of size $N_x \times N_y$ with periodic boundary conditions,
and $\langle i,j\rangle$ denotes the restriction to nearest neighbours. We write $i=(i_x,i_y)$ to access the 2D coordinates
of site $i$.
The hopping integral is $t$, the pairing amplitude $\Delta$,
and the chemical potential $\mu$. The operators $c_{i,\sigma}$ annihilate an electron
of spin $\sigma$ on site $i$, and $c_{i,\sigma}^\dagger$ are the corresponding creation operators.

The interactions with the magnetic impurities have amplitudes $\hat{V}_m$ (denoted differently from
the $V_m$ of the continuum model) and are expressed through the Hamiltonian
\begin{equation}
H_m = \hat{V}_m \sum_{i = (i_x,0)} \mathbf{M}_i \cdot \mathbf{S}_i.
\end{equation}
Here $\mathbf{S}_i = \sum_{\sigma,\sigma'} \boldsymbol{\sigma}_{\sigma,\sigma'} c_{i,\sigma}^\dagger c_{i,\sigma'}$
are the electron spin operator, for $\boldsymbol{\sigma}$ the vector of Pauli matrices, and
$\mathbf{M}_i$ are unit vectors that are either aligned ferromagnetically or wind
in a planar spiral with wave number $k_m$ in the spin $(x,y)$ plane.

For the finite chain in Fig.\ \ref{fig:lowest_eigval} we consider a system of size $N_x=600, N_y=70$
and restrict $H_m$ to values $-224 \le i_x \le 225$ at $i_y=0$. The parameters are chosen such that
$\Delta = 0.1 t, \mu = -3.6 t$ and $a k_F = \arccos[(-\mu-2t)/(2t)] \approx 0.64$.

For the chains with infinite $x$ extension we partially diagonalize the Hamiltonian by performing the
Fourier transform $i_x \to k_x$. For a ferromagnetic alignment ($k_m = 0$) this is done directly.
For spiral magnetizations with $k_m$ we choose the spin axes such that $\mathbf{M}_i$
rotates in the spin-$(x,y)$ plane so that the same gauge transformation
$k_x \to k_x \pm k_m$ as for the continuum model
maps the spiral back to a ferromagnetic alignment.
The periodic boundary conditions along the $x$ directions are always applied in the gauge transformed basis.
Solutions are carried out as described further in Part I.

Green's functions are obtained through the Lehmann representation in terms of the eigenfunctions and eigenvalues of the
Hamiltonian, as a function of $k_x, i_y, i_y'$ and $\omega$.
Topological invariants are calculated by self-consistent determination of the full Hamiltonian, followed by the calculation
of the Pfaffian invariant of the 1D topological Hamiltonians obtained from the inverse of the Green's function at $i_y=i_y'$,
$k_x=0$ and $\omega=0$ in the same way as for the analytic model described in Sec.\ \ref{sec:localized_classification}.
We include only the single time-reversal invariant momentum $k_x = 0$ rather than adding
the influence of the $k_x = \pi/a$ point for better comparison to the continuum model.

In Fig.\ \ref{fig:topol_phasediagram_num} as well as in
the insets of Fig. \ref{fig:topol_phasediagram} the system size is $N_x=51$ and $N_y=100$, and the gap is
self-consistently tuned to $\Delta = 0.1 t$ for $\hat{V}_m=0$. The self-consistent parameters so determined are then used
as input to the diagonalization of the Hamiltonian with added magnetic impurity chain with a variety of $\hat{V}_m$ and $k_m$ values
to determine the phase diagrams. The insets correspond to phase diagrams at sites (a): $i_y = 50$ (i.e. the centre), (b):
$i_y = 51$, (c): $i_y = 52$, (d): $i_y = 58$ and (e): $i_y = 81$. As $a k_F \approx 0.64$ these roughly correspond to the values for
$y k_F$ displayed for the continuum phase diagrams. Note that due to the numerics being on a lattice (b) is as close to the interface
as possible but is not sufficiently close to exactly match the behaviour seen in the continuum model.


\end{document}